\documentclass{aastex631}

\shorttitle{A simulated study of the influence of Martian moons rotation}
\shortauthors{Y. Yang et al.}
\graphicspath{{./}{figures/}}
\usepackage{graphicx,times}
\usepackage{float}
\usepackage{natbib}
\usepackage{amssymb,amsmath}
\usepackage{array}
\begin{document}

\title{A novel ephemeris model for Martian moons incorporating their free rotation}

\author{Yongzhang Yang}
\affiliation{Yunnan Observatories, Chinese Academy of Sciences, Kunming 650216, PR China}

\author{Kai Huang}
\affiliation{Yunnan Observatories, Chinese Academy of Sciences, Kunming 650216, PR China}
\affiliation{University of Chinese Academy of Sciences, Beijing 100049, PR, China}

\author{Jianguo Yan}
\affiliation{State Key Laboratory of Information Engineering in Surveying, Mapping and Remote Sensing, \\
Wuhan University, Wuhan 430070, PR China}


\author{Yuqiang Li}
\affiliation{Yunnan Observatories, Chinese Academy of Sciences, Kunming 650216, PR China}
\email{yang.yongzhang@ynao.ac.cn}



\begin{abstract}

High-precision ephemerides not only support space missions, but can also be used to study the origin and future of celestial bodies. In this paper, a coupled orbit-rotation dynamics model that fully takes into account the rotation of the Martian moons is developed. Phobos and Deimos' rotation are firstly described by Eulerian rotational equations, and integrated simultaneously with the orbital motion equations. Orbital and orientational parameters of Mars satellites were simultaneously obtained by numerical integration for the first time. In order to compare the differences between our newly developed model and the one now used in the ephemerides, we first reproduced and simulated the current model using our own parameters, and then fit it to the IMCCE ephemerides using least-square procedures. The adjustment test simulations show Phobos and Deimos‘ orbital differences between the refined model and the current model is no more than 300~$m$ and 125~$m$, respectively. The orientation parameters are confirmed and the results are in good agreement with the IAU results. Moreover, we simulated two perturbations (main asteroids and mutual torques) which were not included in our refined model, and find that their effects on the orbits are completely negligible. As for the effect on rotation, we propose  to take care of the role of mutual attraction in future models.

\end{abstract}



\section{Introduction}           
\label{sect:intro}

Mars is most similar to the Earth in the Solar System and is the only terrestrial planet other than Earth to have natural satellites. Phobos and Deimos are the two moons of Mars. Since their discovery in 1877, the orbital motion of the Martian moons have been studied extensively. In order to fit the observation data to the study of the dynamical properties of the Martian moons: first Earth-based observations and then to spacecraft observations, a variety of dynamical models have been developed. During the Mariner and Viking era, \citet{sinclair1971motions, sinclair1978orbits} and \citet{shor1975the} employed analytical expressions to fit various sets of Earth-based observations to generate the ephemerides and confirmed the secular tidal acceleration which was firstly studied by \citet{sharpless1945secular}. \citet{jacobson1989orbits} and \citet{sinclair1989orbits} used all available positional observations of satellites of Mars, including Earth-based and from the Mariner 9 and Viking spacecraft, to re-determine the orbits of the Martian moons, these ephemerides are available in the SPICE \citep{arora2010fast} library until now.

The first completely numerical dynamical model of Martian moons were studied by \citet{lainey2007first} during the Mars Express (MEX) mission, the model presented in this work used: $(1)$ the aspherical Martian gravity field, $(2)$ the perturbations of the Sun, Jupiter, Saturn, the Earth, and the Moon using planetary ephemerides, $(3)$ the IAU2000 Martian precession/rotation, $(4)$ the mass of each Martian moon, and the tidal effect was modeled by the tidal bulge raised by each moon on Mars using physical formulation instead of fitting secular accelerations in the satellite longitudes. After fitting to the MEX, Mars Global Surveyor (MGS), Phobos 2, Viking 1-2, Mariner 9, and ground based observations, new ephemerides of the Martian moons have been developed on the basis of the first numerical dynamical model. It is worth mentioning that, although the authors realize that the perturbations due to the librations of the Martian moons have a significant influence that cannot be ignored, this effect was not modeled due to the lack of an accurate $C_{20}$ and estimate libration angles \citep{lainey2007first}.

\citet{jacobson2010orbits} upgraded the dynamical model of the Martian satellites, introducing the effect of Phobos' libration in the form of an analytical formula for the first time. 
Assuming Phobos is rotating synchronously, its pole is perpendicular to the orbital plane, and its axis of minimum principal moment of inertia points toward Mars. The angle between Phobos' axis of minimum principal moment of inertia and the direction from Phobos to Mars also called libration angle is small and can be described by
\begin{equation}
\label{eq:librangle}
\begin{split}
 \theta = (2e + \mathcal{A})\sin M, \,
\end{split}
\end{equation}
 where $\mathcal{A}$ is the libration amplitude and can be calculated from the moment of inertia \citep{chao1989gravitational}, $e$ and $M$ are Phobos' orbital eccentricity and mean anomaly in its orbit, respectively. The revised orbits of Martian moons were obtained by fitting the numerical dynamical model to all available Earth-based observations, imaging observations and radio tracking data from spacecraft.

The above dynamical model has been used until now, and although observational data have been accumulating over time, the dynamical model has introduced only minor additions. Examples include (1) mutual attraction between satellites, general relativity, and (3) the second gravitational field of satellites  \citep{jacobson2014martian, lainey2020mars}. This is mainly due to the difficulty of determining the gravitational coefficients of Phobos and Deimos with current observations.

The Martian Moons eXploration (MMX) mission is under development by the Japan Aerospace Exploration Agency (JAXA), and is scheduled to be launched in 2026. This mission is dedicated to survey the two Martian moons and return samples from Phobos \citep{kawakatsu2023preliminary}. In order to achieve the objective of collecting samples from Phobos, a probe will land on the surface of Phobos. At that time, there will be tracking data between the lander and the Earth, and this new data will offer new opportunities to study the orbital and rotational motions of the Martian moons.

Inspired by MMX mission, this paper presents a refined numerical dynamical model for the Martian moons libration. The libration is described using the Euler-Liouville equations, i.e., with a state of complete free rotation that does not take into account any assumptions. 
This dynamical model will allow us to study the motions of the Martian moons in the future with more realistic scenarios based on new observational data, such as the ones from MMX. In Sect.~\ref{sect:re-model} we briefly introduce the dynamical model now in use. In Sect.~\ref{sect:new-Rot} we detail our optimized libration model of the Martian moons. Sect.~\ref{sect:cmp} provides a detailed comparison of these two models, followed by Sect.~\ref{sect:conclusion} where we summarize and conclude the paper.

\section{REVIEW OF NUMERICAL DYNAMICAL MODEL}
\label{sect:re-model}

In this work, we will use the numerical model now in use as a reference to study its differences with our revised model. Hence, we first introduce the modeling process of an ephemerides model.

The equation of translational motion are described in a planetocentric (Mars) reference system with fixed axes align with the International Celestial Reference System (ICRS). The position vectors of the eight planets and the Sun relative to Mars in this reference system can be easily retrieved from the numerical ephemerides. 
Here we use the latest version of the planetary ephemeris INPOP21a provided by Institut de M{\'e}canique C{\'e}leste et de Calcul des {\'E}ph{\'e}m{\'e}rides (IMCCE) to obtain the position and velocity vectors of the Sun and planets relative to Mars in Barycentric Celestial Reference System (BCRS) \citep{fienga2021inpop21a},.
This ephemeris updates the Mars orbit relative to INPOP19a \citep{fienga2019inpop19a}, adding an additional 2 years data from Mars Express. 

The orbital motion of the satellites around Mars can be described in terms of position ${\bf r} \equiv (\mathrm{r_x}, \mathrm{r_y}, \mathrm{r_z})$ and velocity ${\bf v} \equiv (\mathrm{v_x}, \mathrm{v_y}, \mathrm{v_z})$ in rectangular coordinates. The classical differential equations of relative motion in the planetocentric system can be read as,
\begin{equation}
\label{eq:orb}
\begin{split}
\frac{d^2{\bf r}}{dt^2} = \frac{{\bf F}_\mathrm{s}}{\mathrm{m_s}} - \frac{{\bf F}_\mathrm{0}}{\mathrm{m_0}}\,,
\end{split}
\end{equation}  
where ${\bf F}_\mathrm{s}$ and ${\bf F}_\mathrm{0}$ indicate the whole external forces exerted on the satellites and Mars, $\mathrm{m_s}$ is the mass of satellite, $\mathrm{m_0}$ is the mass of Mars, and $t$ is the time expressed in Barycentric Dynamical Time (TDB) timescale.
 
The forces that induce the relative motion can be split up into a two-body part, a part for third-body perturbation, a part for mutual attraction, a part for tidal perturbation, a part for relativistic perturbation and a part for spin librations. Hence, the $\frac{d^2{\bf r}}{dt^2}$ can be rewritten as :
\begin{equation}
\label{eq:acc}
\begin{split}
\frac{d^2{\bf r}}{dt^2} = {\bf a}_\mathrm{two-body} +  {\bf a}_\mathrm{third-body} + {\bf a}_\mathrm{mutual}  +{\bf a}_\mathrm{tide}  +{\bf a}_\mathrm{rel} +{\bf a}_\mathrm{libr}\,.
\end{split}
\end{equation} 
Here the ${\bf a}_\mathrm{two-body}$,  ${\bf a}_\mathrm{third-body}$ and $ {\bf a}_\mathrm{rel}$ have the usual form used in numerical ephemerides \citep{viswanathan2017inpop17a, pitjeva2017epm2017, folkner2014planetary}.

If we consider Deimos as a third body, we can calculate the effect of Deimos on Phobos' orbit. According to the third-body perturbation equation, the acceleration of Phobos due to mutual attraction can be described as:
\begin{equation}
\label{eq:attrp}
\begin{split}
{\bf a}_\mathrm{mutual,p} =  \mathrm{\mu_d} \left( \frac{\bf{r_d- r_p}}{\mathrm{r^3_{pd}}} - \frac{\bf{r_d }}{\mathrm{r^3_{d}}} \right)\,,
\end{split}
\end{equation} 
Similarly, the acceleration of Deimos due to mutual attraction is:
\begin{equation}
\label{eq:attrd}
\begin{split}
{\bf a}_\mathrm{mutual,d} =  \mathrm{\mu_p} \left( \frac{\bf{r_p- r_d}}{\mathrm{r^3_{dp}}} - \frac{\bf{r_p }}{\mathrm{r^3_{p}}} \right)\,.
\end{split}
\end{equation} 
Where $\mu_d$, $\mu_p$, $\bf{r_d}$ and $\bf{r_p}$ are the product of gravitational constant and mass of the Deimos and Phobos, the position vectors of Deimos and Phobos relative to Mars in the inertia system, respectively. $\mathrm{r_p}$ denotes the norm of $\bf{r_p}$,  $\mathrm{r_d}$ denotes the norm of $\bf{r_d}$, and $\mathrm{r_{dp}} = \mathrm{r_{pd}}=|\bf{r_p - r_d}|$.

For tidal acceleration ${\bf a}_\mathrm{tide}$, \citet{lainey2007first} and \citet{jacobson2010orbits} employed a different but similar form of model in their numerical integration. In this paper, we simulate the differences between our new model and the French Numerical Orbit and Ephemerides (NOE), hence, here we refer to \citet{lainey2007first} for a complete description of the tidal force ${\bf F}_T$ acting on the satellite of the form

\begin{gather}
\label{eq:stide}
{\bf F}_T = -\frac{3k_2\mathrm{\mu_s}\mathrm{m_s}a^{5}_{m}}{\mathrm{r^8}} \left(  
{\bf r} +\Delta t \left[
\frac{2{\bf r}({\bf r} \cdot {\bf v})}{\mathrm{r^2}} + ({\bf r} \times \bf \Omega + \bf v)
\right]
\right) \,,
\intertext{where:}
  \begin{tabular}{>{$}r<{$}@{\ =\ }l}
    k_2 & the Love number of Mars\\
    \bf \Omega & the Martian angular velocity vector \\
    \mathrm{\mu_s} & $GM$ of Satellite \\
    \mathrm{m_s} & mass of Satellite \\
    \Delta t & the time delay due to viscoelastic response of Mars \\
    \mathrm{a_m} & the equatorial radius of Mars. \\
  \end{tabular}\nonumber
\end{gather}

Finally, we present the satellites' figure acceleration and libration model used in the current ephemerides of Martian moons. 
Under the assumption that the spin pole is normal to the orbital plane, according to Eqn.~\ref{eq:librangle}, The quadrupole force on Mars exerted by satellite can be computed as:
\begin{equation}
\label{eq:slib}
\begin{split}
{\bf F}_m =  \frac{3}{2}{\mu_s}
\left (
\frac{{a_m}^2}{r^4}
\right )
[(-C_{20} + 6 C_{22} \cos2\theta) ~\hat{\bf{r}} + 4 C_{22} \sin2\theta ~\hat{\bf{t}}] \,,
\end{split}
\end{equation} 
where $C_{20}$ and $C_{22}$ are the second zonal and sectorial harmonic of satellite, $\hat{\bf{r}}$ denotes the unit vector directed from Mars toward satellite and $\hat{\bf{t}}$ denotes the unit vector in the satellite's orbit plane normal to $\hat{\bf{r}}$ and in the direction of its orbital motion. Therefore, the reaction force acting on the satellite is
\begin{equation}
\label{eq:slib2}
\begin{split}
{\bf F}_s = -
\left(
\frac{\mu_m}{\mu_s}
\right)
~{\bf F}_m
 \,,
\end{split}
\end{equation} 
with $\mu_m$ denotes the $GM$ of Mars.

Use these formulas Eqs.~\ref{eq:attrp}~-~\ref{eq:slib2} to calculate ${\bf a}_\mathrm{mutual}$, ${\bf a}_\mathrm{tide}$, and ${\bf a}_\mathrm{libr}$ in turn and summing them up together with ${\bf a}_\mathrm{two-body}$,  ${\bf a}_\mathrm{third-body}$ and $ {\bf a}_\mathrm{rel}$, we then have the orbital equations of motion in the planetocentric coordinates.

\section{Rotation model of the Martian moons}
\label{sect:new-Rot}
In this paper, the Martian moons are modeled as rigid bodies as in our previous study of Phobos' libration \citep{yang2020elastic}. The orientation of Phobos and Deimos are integrated from the differential equations for their angular velocities. The angular momentum vector of a satellite is the product of angular velocity and moment of inertia. The angular momentum vectors varies with time due to external torques.

The process of modeling rotation is presented here using Phobos as an example. In order to describe the variations of Phobos' rotation in the inertial frame, for convenience, the frame of Phobos was aligned with its principal axis (PA). Then the orientation of the Phobos‘ frame with respect to the inertial frame is determined by three Euler angles: $\phi$, $\theta$, and $\psi$, which vary with time $t$. The transformation from the inertial system to the body-fixed system (PA) is given by the matrix:
\begin{equation} 
\label{eq:rotm}
R_{B2C}= R_{z}(\phi(t))R_{x}(\theta(t))R_{z}(\psi(t))\,,
\end{equation}
where the rotation matrices $R_{z}$ and $R_{x}$ are right-handed rotations around $z$-axis and $x$-axis, respectively. Hereafter, for simplicity, the argument $t$ will be omitted where appropriate.

In a rotating system, the rate of change of angular velocity $\bm{\omega}$ is related to the torque $\bf{T}$ and determined by Euler-Liouville's equations of rotation,
\begin{equation} 
\frac{d}{dt}(\mathbf{I}{\bm \omega}) + \bm{\omega} \times {\mathbf{I}\bm{\omega}} = {\bf T} \,,
\label{eq:eul2}
\end{equation}
where $\mathbf I$ represents the moment of inertia tensor. This leads to the equations for $\bm{\omega}$, consequently, 
\begin{equation}
\begin{split}
\frac{d{\bm{\omega}} }{dt}= \mathbf{I}^{-1}
\left(
 \mathbf{T} - 
\frac{d\mathbf{I}}{dt} \bm{\omega} - \bm{\omega} \times \mathbf{I} \bm{\omega} 
\right)\,.
\end{split}
\label{eq:eul3}
\end{equation}
The effect of elastic deformation is not considered in the current modeling because its effect on the Phobos ephemeris is less than 100~$m$ even if Phobos $k_2$ is as large as $1\times10^{-4}$ \citep{yang2024numerical}, corresponding to extremely porous Phobos \citep{le2013phobos}, which we consider unlikely.
Since the porosity of Phobos is anyway limited by the density of the material that makes the matrix of the bulk material. 
Hence, $\mathbf{I}$ is diagonal and constant. Solving Eqn.~\ref{eq:eul3} for $\frac{d{\bm{\omega}} }{dt}$ we find that the resultant angular acceleration takes a simple form:
\begin{equation}
\begin{split}
\frac{d{\bm{\omega}} }{dt}= \mathbf{I}^{-1}
\left(
 \mathbf{T} - \bm{\omega} \times \mathbf{I} \bm{\omega} 
\right)\,.
\end{split}
\label{eq:eul4}
\end{equation}
The components of the angular velocity vector in the body-fixed system are easily expressed in terms of reference Euler angles $(\phi, \theta,\psi)$ \citep{goldstein2002classical}:
\begin{eqnarray}
\begin{aligned}
\omega_{1} &= \dot{\theta}\cos\psi + \dot{\phi}\sin\psi\sin\theta   \\
\omega_{2} &= -\dot{\theta}\sin\psi + \dot{\phi}\cos\psi\sin\theta   \\
\omega_{3} &= \dot{\phi}\cos\theta + \dot{\psi}\,,
\end{aligned}
\label{eq:eul1}
\end{eqnarray}
where $(\phi, \theta,\psi)$ are the precession angle, nutation angle, and rotation angle, respectively.

If we differentiate the Eqns.~\ref{eq:eul1} with respect to time $t$ and rearrange it, we get a linear system of equations containing $\ddot{\phi}$, $\ddot\theta$ and $\ddot\psi$,

\begin{eqnarray}
\begin{aligned}
 \ddot{\phi}   &= \csc\theta(\dot{\omega}_{1}\sin\psi + \dot{\omega}_{2}\cos\psi + \dot{\theta}\dot{\phi}) - \dot{\psi}\dot{\theta}\cot\theta  \\
 \ddot{\theta} &= \dot{\omega}_{1}\cos\psi -  \dot{\omega}_{2}\sin\psi - \dot{\phi}\dot{\psi}\sin\theta  \\
 \ddot{\psi}    &= \dot{\omega}_{3} - \ddot{\phi}\cos\theta + \dot{\psi}\dot{\theta}\sin\theta \,.
\end{aligned}
\label{eqn:eul5}
\end{eqnarray}
The above equations model the Euler angle equations of motion, the key is to calculate the angular acceleration $\frac{d{\bm{\omega}} }{dt}$, which can be evaluated by Eqns.~\ref{eq:eul2} - \ref{eq:eul4} . Having established the mathematical equation for the Euler angle and external torques through angular acceleration, we now derive the calculation of external torques.

The calculation of the torque $\bf T$ applied to a satellite is usually divided into two parts:  a torque from point-mass (body) A to the satellite's figure and a torque from the oblateness ($J_2$) of Mars to the satellite’s figure,
\begin{equation}
\begin{split}
\mathbf{T} = \mathbf{T}_{pm} +  \mathbf{T}_{fig}  \,,
\end{split}
\label{eq:12}
\end{equation}
A detailed description of $\bf T$ and $\bf I$ can be found at \citet{williams2001lunar, rambaux2012rotational, folkner2014planetary}, \citet{pavlov2016determining} and \citet{yang2020elastic}. 

The instantaneous state of rotation of a rigid body can be defined completely by the six quantities, i.e. the above-defined Euler angles and their rates of change. In this paper, we use the above approach to model the libration of Phobos and Deimos, unlike the models used in the present ephemerides (pole normal to its orbit plane), the Martian moons are completely free to rotate without any assumptions.

\section{Comparison}
\label{sect:cmp}
The studies in the rotation and orbital motion of the Martian moons described in the previous sections are motivated by the high-precision observational data that may be available from future missions, such as the prober's orbital data and satellites image data when the prober is at a very close distance. In addition, in particular landing and lander tracking measurements carried out on Phobos \citep{usui2018martian, kawakatsu2017mission}. In this section, we simulate our new dynamical model incorporating the free rotation of the Martian moons, and then compare and analyze our model with respect to the current ephemeris models. 

\subsection{Methodology}
\label{sect:method}
In our numerical model, there are a number of parameters whose values affect the orbit significantly, such as the Martian gravitational field, the satellite's initial position and velocity. In order to clarify the differences between the fully coupled approach and the simple libration model (i.e. Eqns.~\ref{eq:slib} -\ref{eq:slib2}) used so far, we borrowed from previous approaches \citep{yang2024numerical} and first simulated the current simple model \citep{lainey2007first, jacobson2010orbits, jacobson2014martian, lainey2020mars}, but with our own selected physical parameters listed in Table.~\ref{tbl.1}. We then fitted the twelve initial parameters (positions and velocities of Phobos and Deimos) as our solve-for parameters to fit the current ephemeris. This fit resolves the issue of differing parameters and provides the best for reference to investigate the differences between the new full model and the ephemeris model used so far. To fit the parameters of the model, we introduce this common relational formula,

\begin{equation}
\begin{split}
\frac{\partial }{\partial P_{j}}\left( \frac{d^2\mathbf{r}_{i}}{dt^2} \right) =
 \frac{1}{m_{i}} \left[
 \sum_{k}\left(
 \frac{\partial{\mathbf{F}_k}}{\partial \mathbf{r}_k}\frac{\partial \mathbf{r}_k}{\partial P_{j}} +
 \frac{\partial{\mathbf{F}_k}}{\partial \dot{\mathbf{r}}_k}\frac{\partial \dot{\mathbf{r}}_k}{\partial P_{j}} +
 \frac{\partial{\mathbf{F}_k}}{\partial \Phi_k}\frac{\partial \Phi_k}{\partial P_{j}} +
 \frac{\partial{\mathbf{F}_k}}{\partial \dot{\Phi}_k}\frac{\partial \dot{\Phi}_k}{\partial P_{j}} +
 \frac{\partial \mathbf{F}_k}{\partial P_{j}}
 \right)
 \right] \,,
\end{split}
\label{eq:var}
\end{equation}
where $\Phi$ and $\dot\Phi$ are the Euler angles and their rates, respectively. In particular, $\Phi$ and $\dot\Phi$ are not modeled in ephemerides model and should be omitted. $P_j$ denoting the unspecified parameter of the model that shall be fit (such as initial  positions, velocities, Euler angles, etc.), then the variational equations are integrated simultaneously with the dynamical model.

\begin{table*}[!htbp]
\small
\caption{Parameters used in dynamical model.} 
\label{tbl.1}
\centering
\begin{tabular}{lcccccc}
\hline\hline
Parameters && Value&Notes &Reference&\\
\hline
The Sun and planets&&$-$&INPOP21a&\citet{fienga2021inpop21a}&\\
Gravity field of Mars&&$-$&MRO120F, up to degree 12&\citet{konopliv2020detection}&\\
Martian precession/rotation&&$-$&Quoted from MRO120F&\citet{konopliv2020detection}&\\
$k_{2,M}$&&$0.169$&Martian love number&\citet{konopliv2020detection}&\\
$Q_{M}$&&$99.5$&Martian dissipation factor&\citet{jacobson2014martian}&\\
$\delta \bar{J}_{3,M}$&&$-$&Seasonal gravity change of Mars&\citet{konopliv2006global,konopliv2011mars,konopliv2020detection}&\\
Gravity field of Phobos&&$-$&Forward model, up to degree 2&\citet{yang2020elastic}&\\
$\mathcal{R}_{P}$& &$10.993$ & Radius,km & \citet{willner2014phobos} &\\
$GM_{P}$& &$0.7072  \times 10^{-3} $&  $GM$ of Phobos& \citet{patzold2014phobos}&\\
Moment of inertia of Phobos&&$0.35545, 0.41811, 0.49134$& Normalized by $MR^2$&\citet{yang2020elastic}&\\
Gravity field of Deimos&&$-$&Forward model, up to degree 2&\citet{rubincam1995gravitational}&\\
$\mathcal{R}_{D}$& &$6.25$ & Radius,km & \citet{rubincam1995gravitational} &\\
$GM_{D}$& &$0.98  \times 10^{-4} $&  $GM$ of Deimos& \citet{konopliv2006global}&\\
Moment of inertia of Deimos&&$0.338, 0.461, 0.508$& Normalized by $MR^2$&\citet{rubincam1995gravitational}&\\
\hline
\end{tabular}
\end{table*}

\subsection{Fit to NOE Ephemerides}
\label{sect:fit2noe}
\begin{figure} [!htbp]
   \centering
   \includegraphics[width=12.0cm, angle=0]{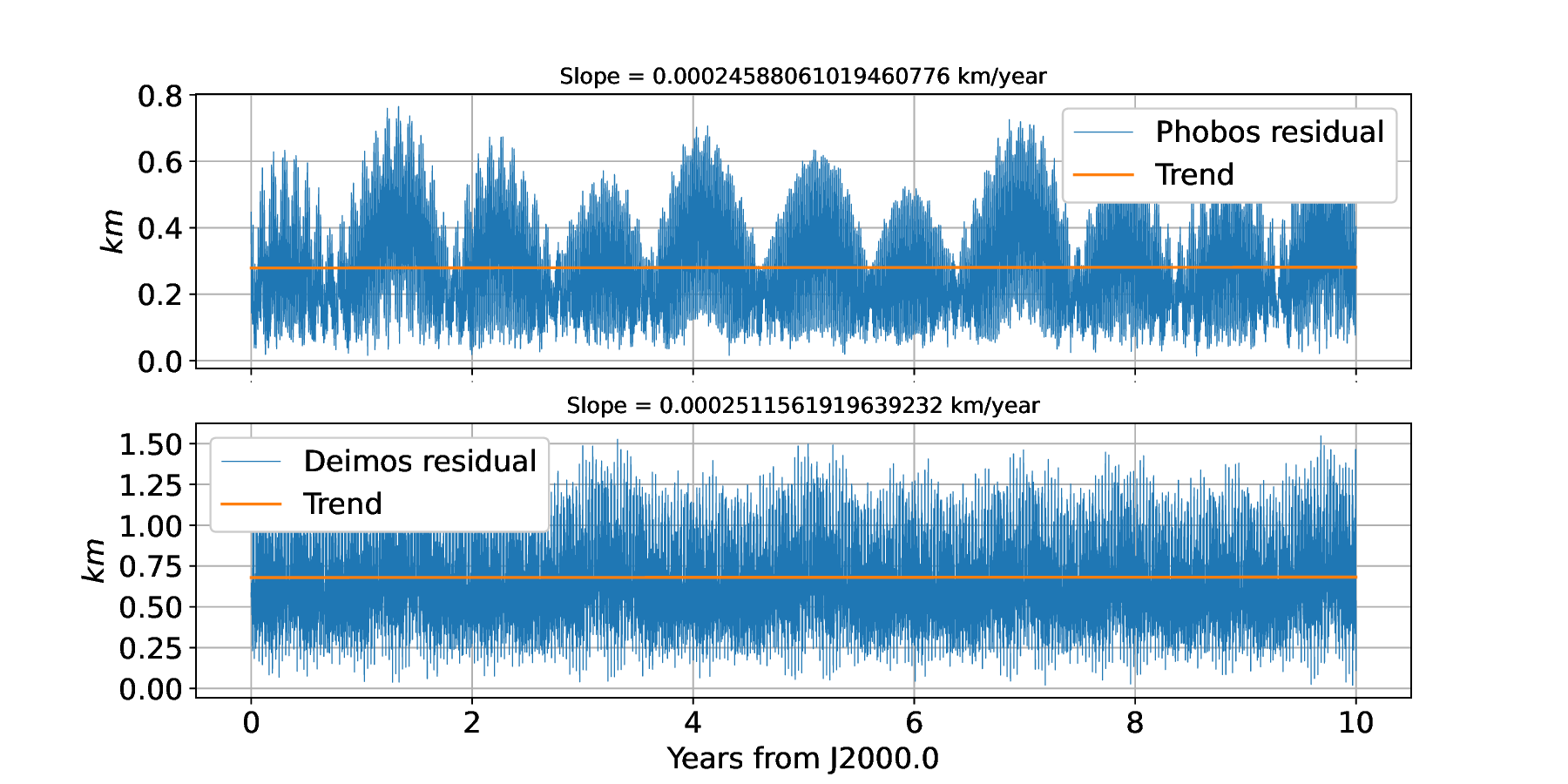}
   \caption{Differences in position after fitting the numerical model to the NOE ephemerides for Phobos (upper panel) and Deimos (lower panel). The satellites' initial positions and velocities have been determined.} 
   \label{Fig1}
   \end{figure}

We chose the latest Martian ephemerides NOE-4-2020 \citep{lainey2020mars} as the "observational data" and fit our simple model to them. The adjustment was performed by least-squares in Cartesian planetocentric coordinates J2000, using a sample set of 3650 points with a step size of one day (ten years), with no weights assigned. We started with the initial epoch at JD 2451545.0 (J2000.0, TDB timescale) and we integrated the model over a decade. The residuals after fit are shown in Fig.~\ref{Fig1} . The resulting difference on the positions are probably explained by the different physical parameters (such as the physical libration $\mathcal{A}$, the $C_{20}$ and $C_{22}$ of the satellites and the Martian dissipation factor Q, etc.) in these two models. 

\subsection{Fit to the ephemeris model}
\label{sect:fit2ref}
\begin{figure} [!htbp]
   \centering
   \includegraphics[width=12.0cm, angle=0]{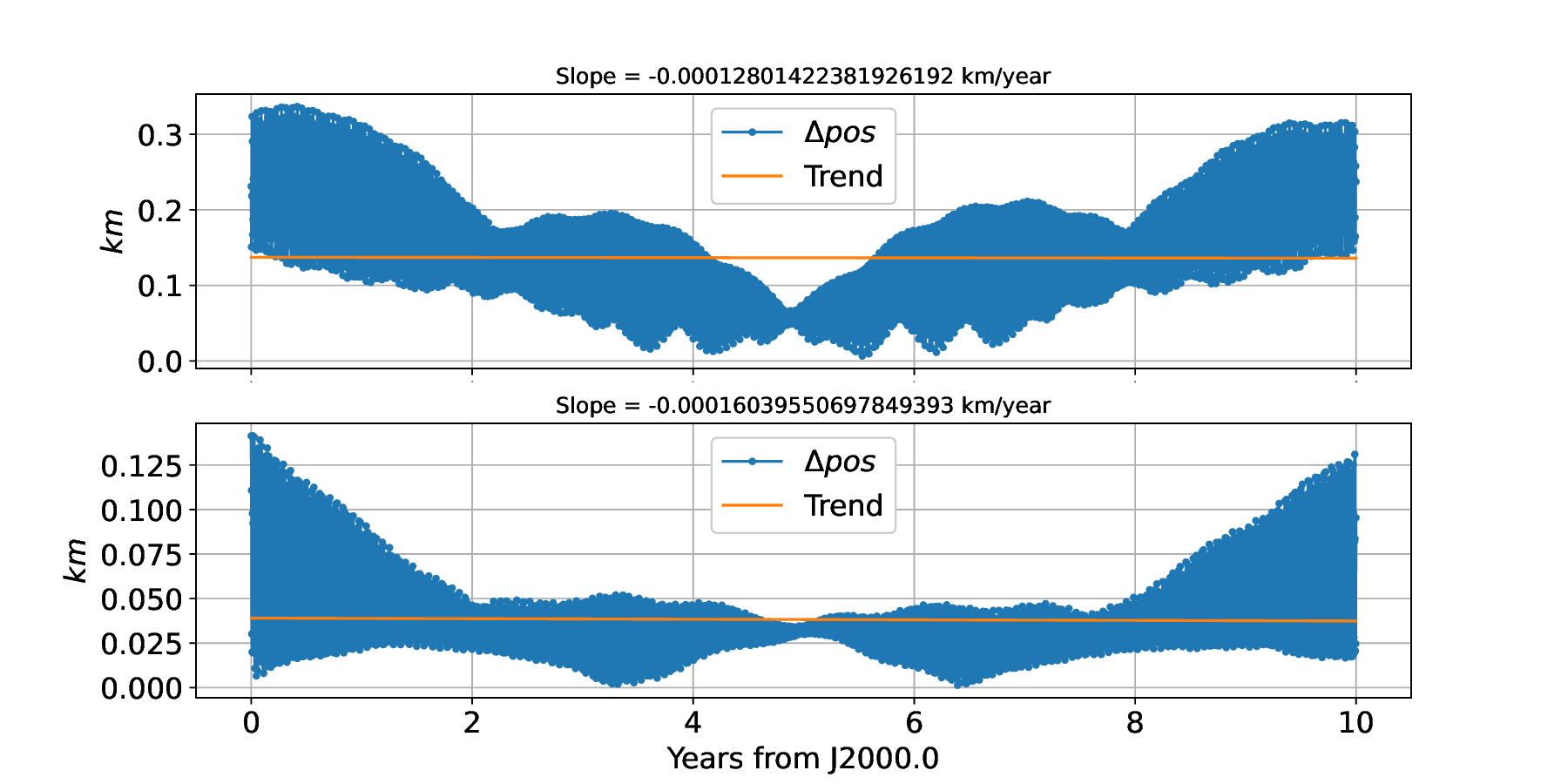}
   \caption{Difference in distances after 10 years’ fitting between our refined model to the simulated simple model, Phobos (upper panel) and Deimos (lower panel). The satellites' initial positions, velocities, Euler angles and its rates have been determined here.} 
   \label{Fig2}
   \end{figure}

In order to explore the differences between our refined new dynamical model and the previous ephemeris model \citep{lainey2020mars, jacobson2014martian, jacobson2010orbits}, we adjusted our new refined model to the simulated ephemeris integrated in Sect.~\ref{sect:fit2noe}, the parameters of these two physical models being identical, so the differences are mainly due to the fact that the details of the models are not identical. We select the twenty-four initial conditions (each satellite's position, velocity, Euler angles and their rates) as our solve-for parameters to fit the simulated result in previous section. The initial Euler angles and rates are referred to the International Astronomical Union (IAU) rotational elements \citep{archinal2018report}. Then the least-square procedures are applied to the model. Fig.~\ref{Fig2} shows the differences in distance after adjustment. The positional deviation is may be due to the newly introduced Phobos latitudinal libration and the different longitudinal librations in the refined model.

\begin{figure} [!htbp]
   \centering
   \includegraphics[width=12.0cm, angle=0]{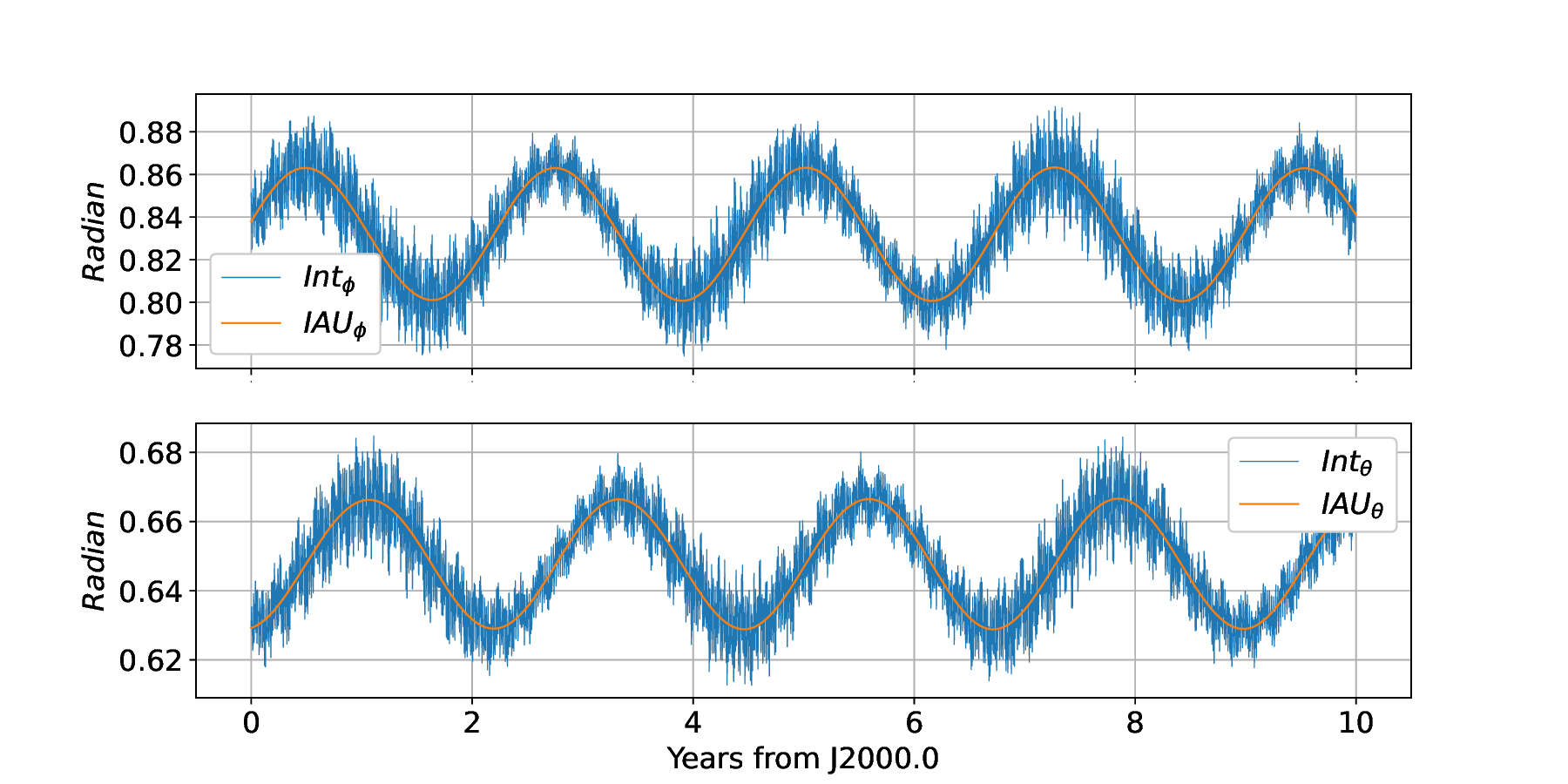}
   \caption{Temporal evolution of Phobos’ precession angle $\phi$ and nutation angle $\theta$ over ten years. Here the numerical integration results are shown in blue and the IAU results are shown in red.} 
   \label{Fig3}
   \end{figure}

\begin{figure} [H]
   \centering
   \includegraphics[width=12.0cm, angle=0]{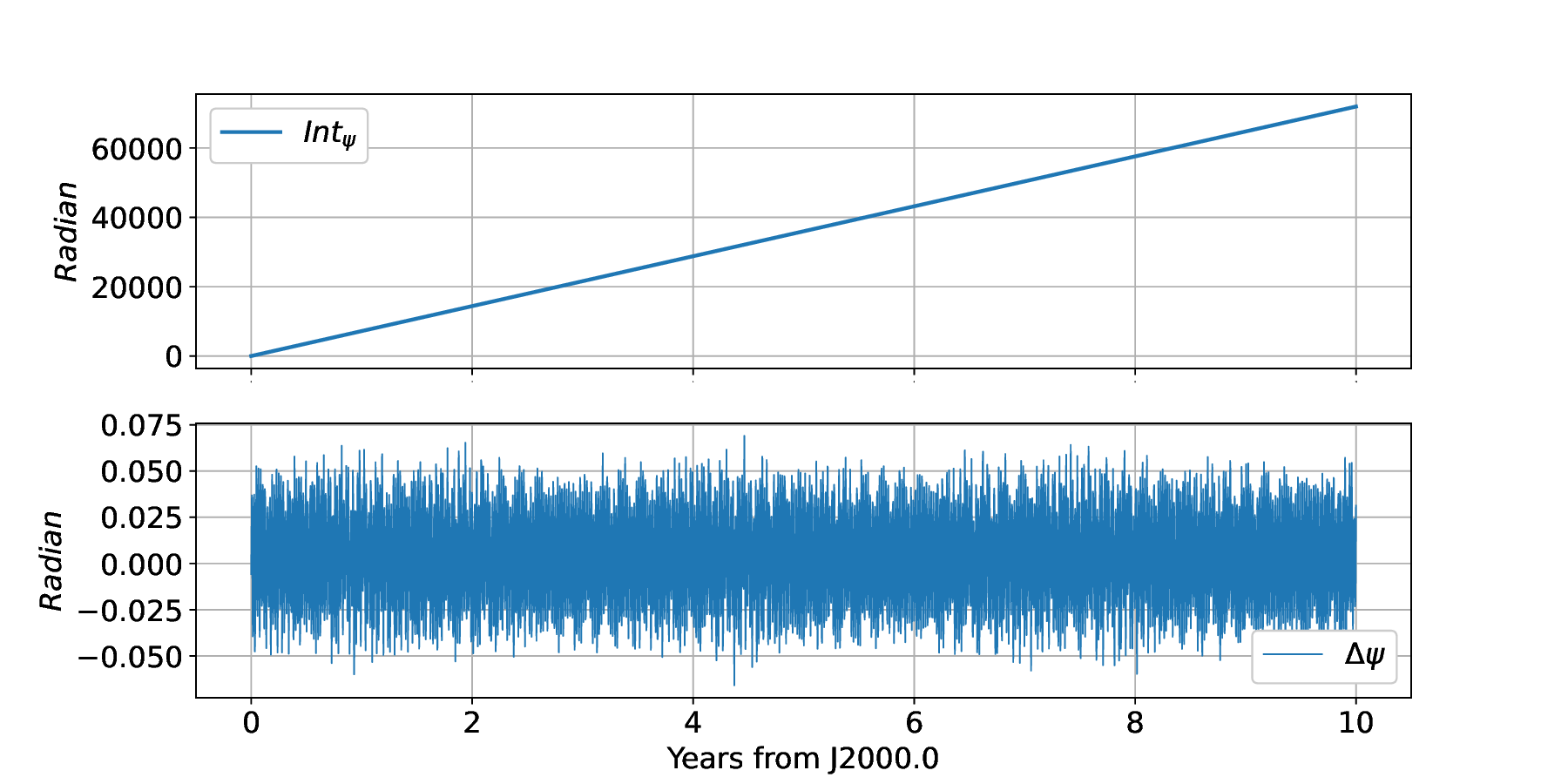}
   \caption{Temporal evolution of Phobos’ rotation angle $\psi$ (upper panel) and its difference to IAU's model (lower panel) about 10 years.} 
   \label{Fig4}
   \end{figure}

\begin{figure} [H]
   \centering
   \includegraphics[width=12.0cm, angle=0]{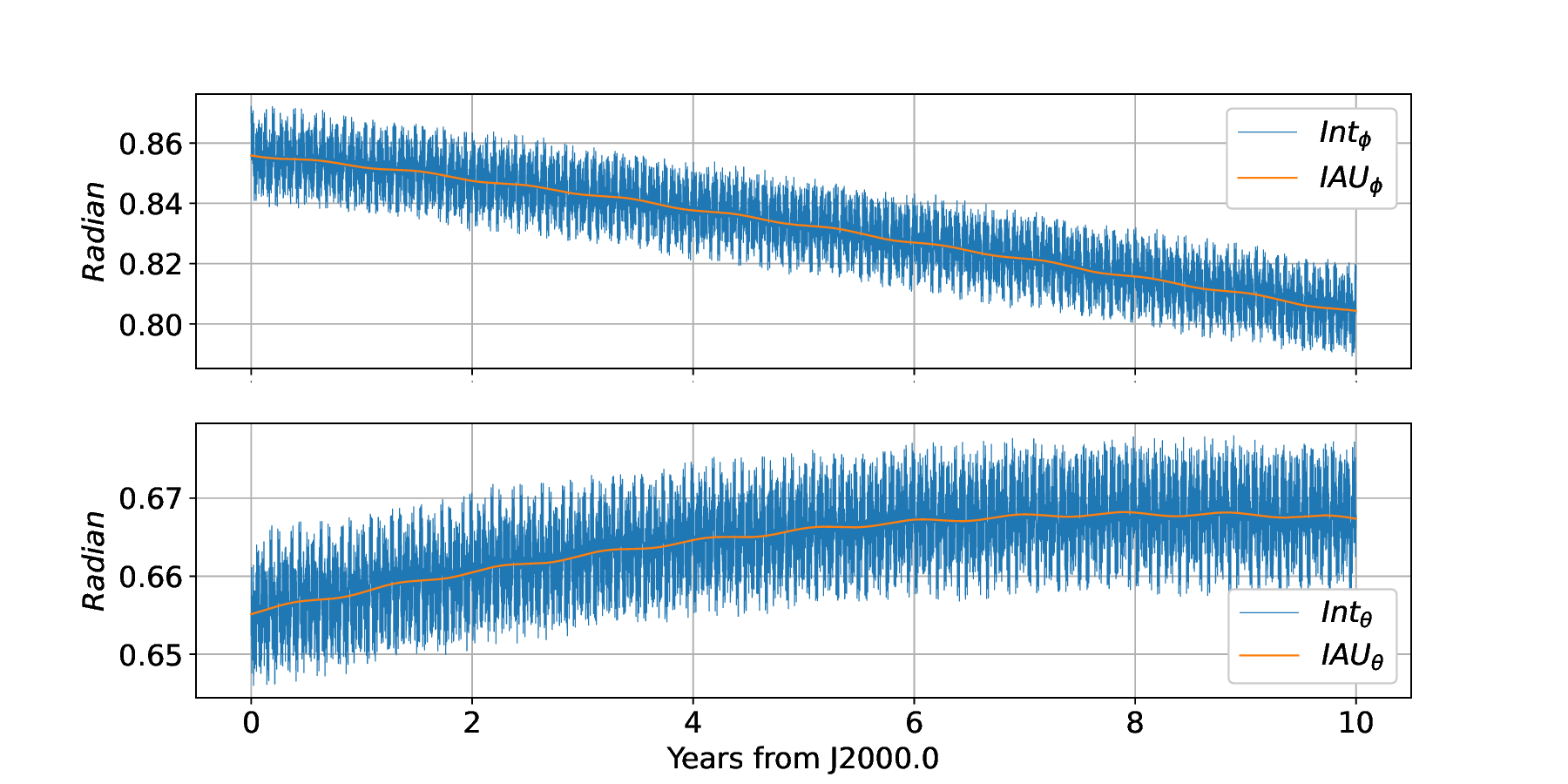}
   \caption{Temporal evolution of Deimos’ precession angle $\phi$ and nutation angle $\theta$ over ten years. The numerical integration results are shown in blue and the IAU models are shown in red.} 
   \label{Fig5}
   \end{figure}

\begin{figure} [H]
   \centering
   \includegraphics[width=12.0cm, angle=0]{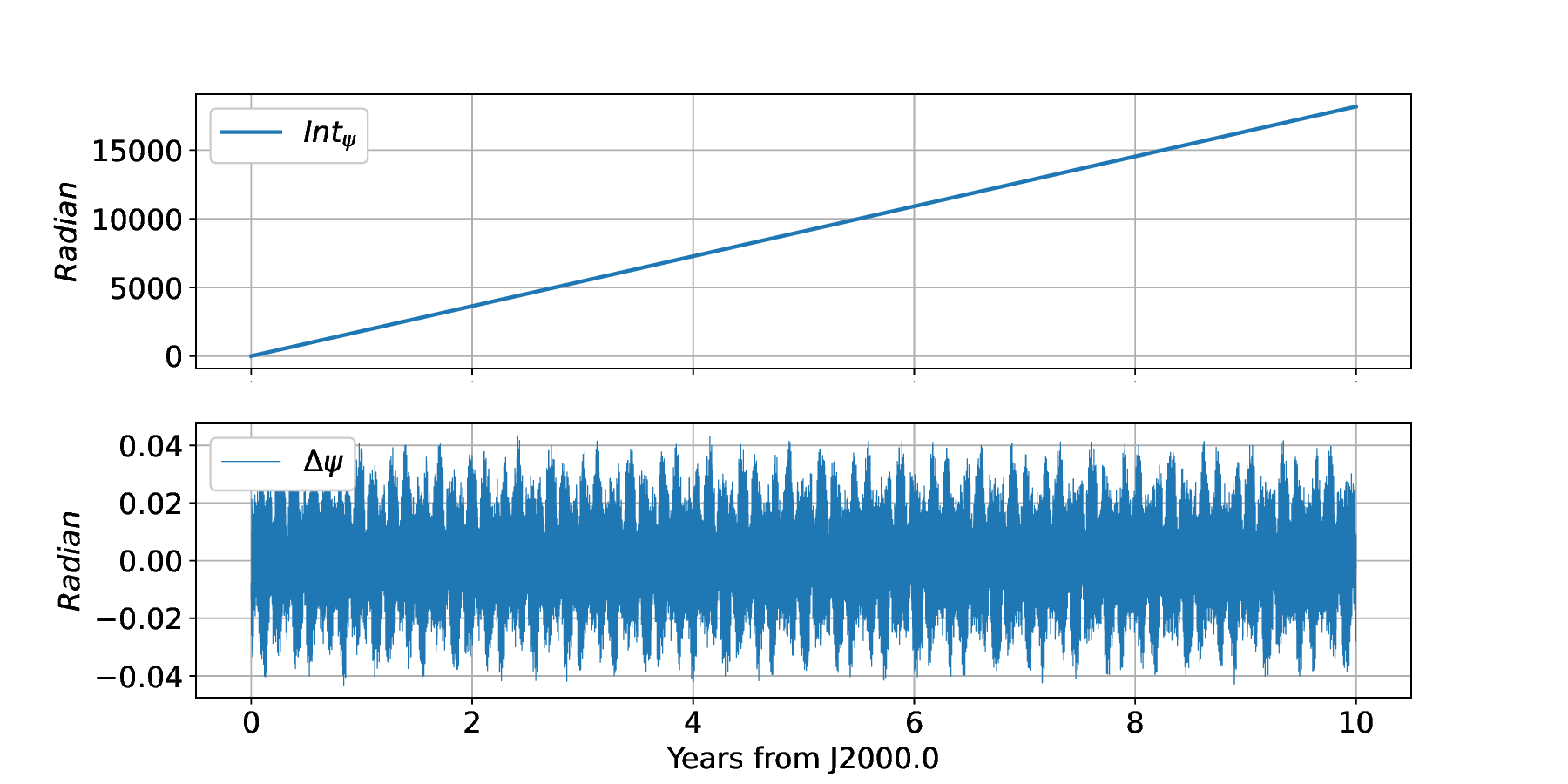}
   \caption{Temporal evolution of Deimos’ rotation angle $\psi$ (upper panel) and its difference to IAU's result (lower panel) over 10 years.} 
   \label{Fig6}
   \end{figure}

 One of the advantages of our refined model is that the Euler angles of the Martian moons and their rates can be obtained by numerical integration.
 Here the evolution of the Euler angles defined w.r.t the inertial reference frame for 10 years from 1 Jan 2000 to 2010 (TDB timescale) are plotted in Figs.~\ref{Fig3} -~\ref{Fig6}. Phobos and Deimos' precession angle $\phi$ and nutation angle $\theta$ are plotted along with IAU's modeling results in Fig.~\ref{Fig3} and Fig.~\ref{Fig5}, respectively. The rotation angles $\psi$ obtained by integration and its difference from the IAU's result are showed in Fig.~\ref{Fig4} and Fig.~\ref{Fig6}, respectively. Although the reference frame used by IAU is slightly different from the one we used \citep{archinal2018report}, the results are still in pretty good agreement. To characterize the high-frequency spectrum of the difference between our numerical integration results and the IAU polynomials, we decomposed the difference in frequency domain by employing the method used in \citet{yang2017determination, yang2019estimation}. The period and frequency of the largest amplitude term is shown in Table.~\ref{tbl.2}. These high-frequency oscillations are due to the rotational motion of the satellites and their orbital motion around the Mars.
 
\begin{table*}[!htbp]
\begin{minipage}[]{100mm}
\caption[]{Frequency analysis of the difference between numerical integration results and the IAU polynomials.\label{tbl.2}}\end{minipage}
\setlength{\tabcolsep}{1pt}
\small
\centering
 \begin{tabular}{ccccccccccccc}
  \hline\noalign{\smallskip}
Satellite&& Arg &&Per(days) && Fre(rad/day)&&Amp(rad)&\\
  \hline\noalign{\smallskip}
 &&$\Delta\phi$ && 0.2333&& 26.9325 && 0.0082&\\
 Phobos&&$\Delta\theta$ && 0.2333&& 26.9325 && 0.0049&\\
 &&$\Delta\psi$ && 0.3190&& 19.6944 &&  0.0311&\\
 \noalign{\smallskip}\hline
  &&$\Delta\phi$ && 0.9596&& 6.5480 && 0.0061&\\
 Deimos&&$\Delta\theta$ && 0.9596&& 6.5480 && 0.0040&\\
 &&$\Delta\psi$ && 1.2625&& 4.9768 &&  0.0281&\\
\noalign{\smallskip}\hline
\end{tabular}
\end{table*}

\begin{figure} [H]
   \centering
   \includegraphics[width=12.0cm, angle=0]{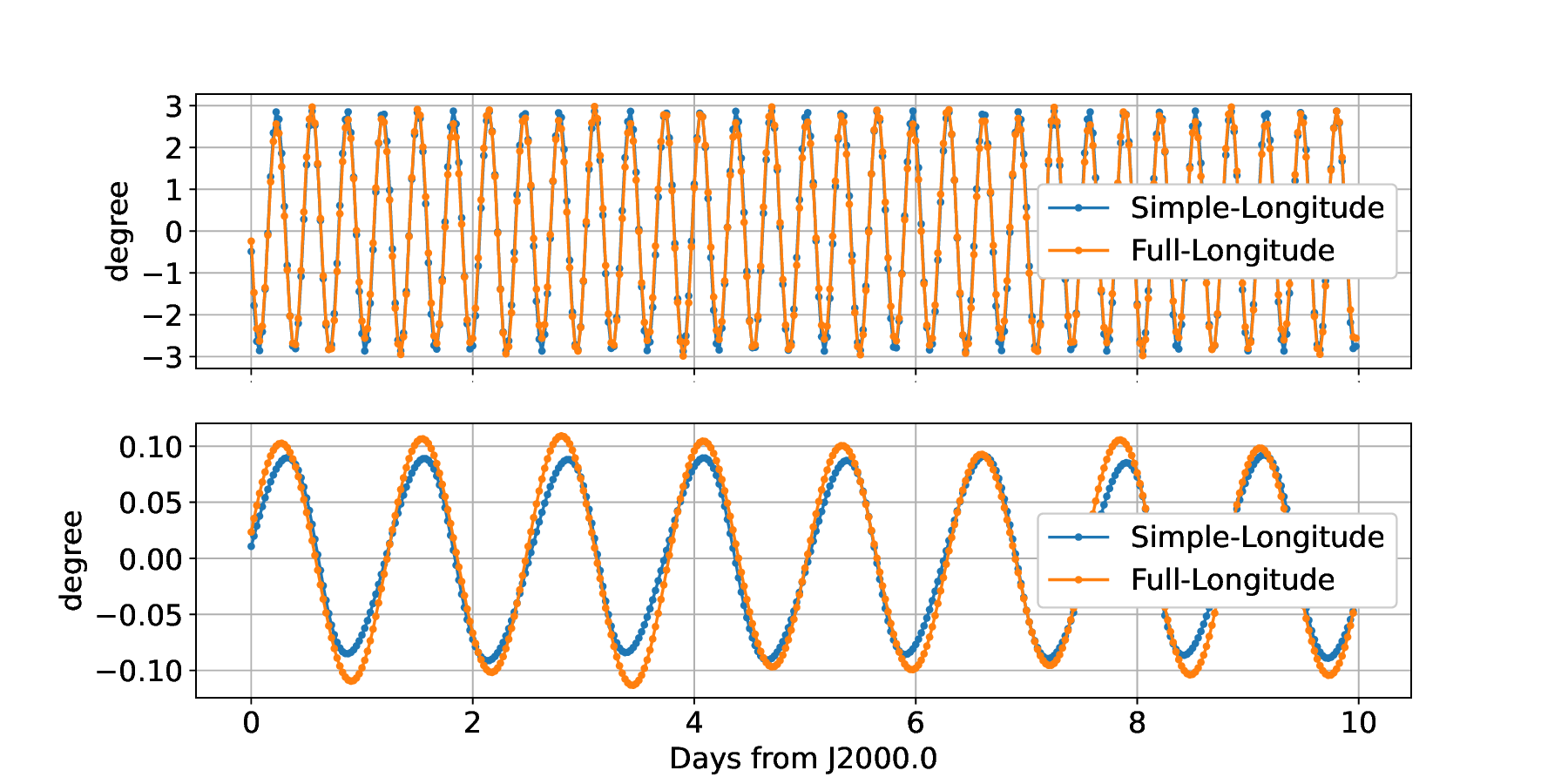}
   \caption{Longitude of Mars in the martian moons' principal-axis coordinate system, upper panel for Phobos, lower panel for Deimos.} 
   \label{Fig7}
   \end{figure}

\begin{figure} [H]
   \centering
   \includegraphics[width=12.0cm, angle=0]{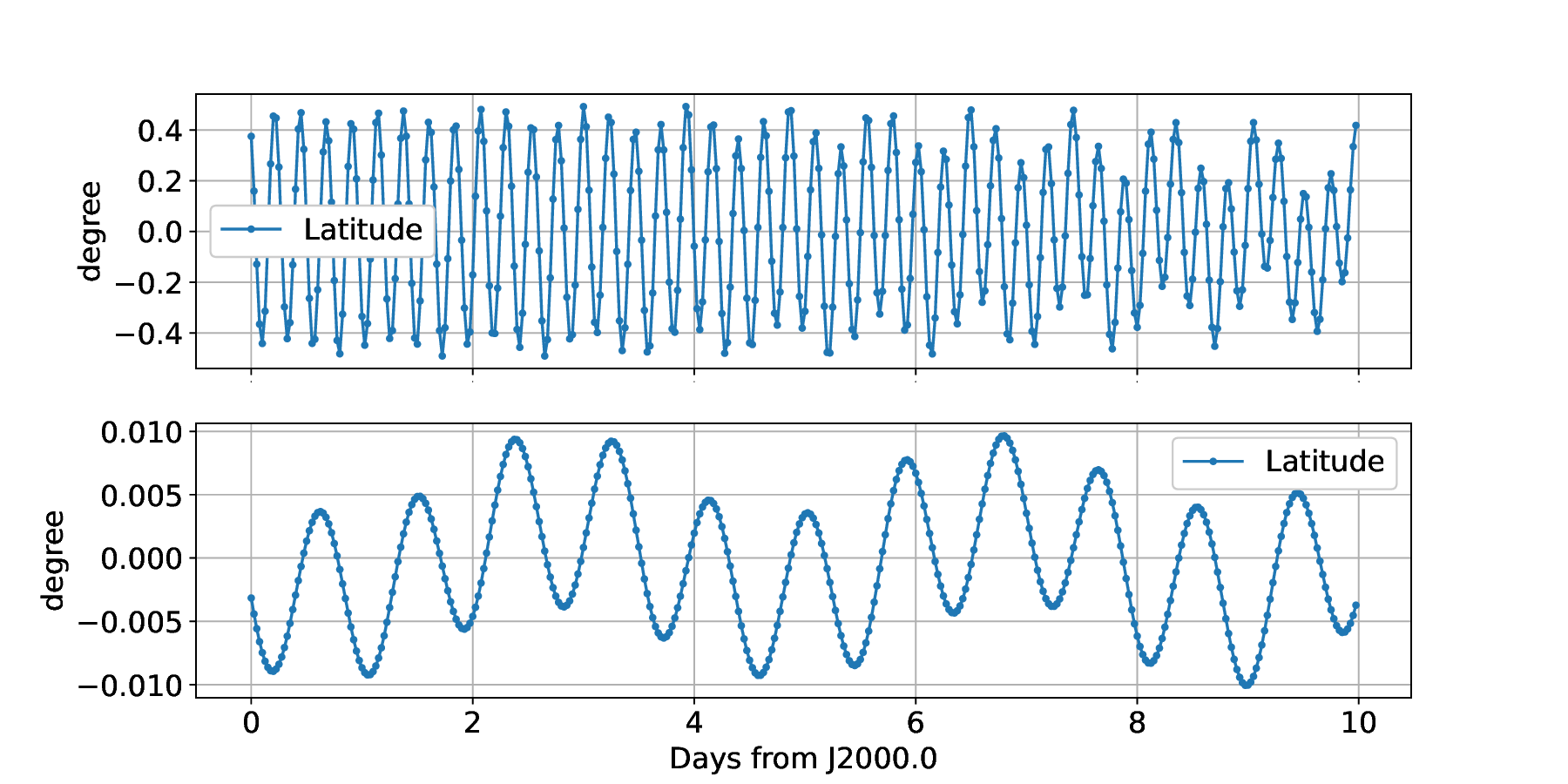}
   \caption{Latitude of Mars in the martian moons' principal-axis coordinate system, upper panel for Phobos, lower panel for Deimos.} 
   \label{Fig8}
   \end{figure}

 To demonstrate the difference between the libration of the two models, Fig.~\ref{Fig7} presents the longitude of the direction from Martian moons to Mars and the moons' axis of minimum principal moment of inertia with respect to the two models, respectively. By comparison we can see that the differences between the two dynamical models are very small, indicating that the deviation in the longitude direction can be described by the simple model well (Eqn.~\ref{eq:librangle}). The moons' obliquities are shown in Fig.~\ref{Fig8}, based on the assumption that the pole is normal to the orbital plane, these parameters were not considered in the simple model, and these values are an order of magnitude smaller than the longitudinal librations.


\subsection{Consideration of two minor perturbations}
\label{sect:minor_perturbation}
\begin{figure} [!htbp]
   \centering
   \includegraphics[width=12.0cm, angle=0]{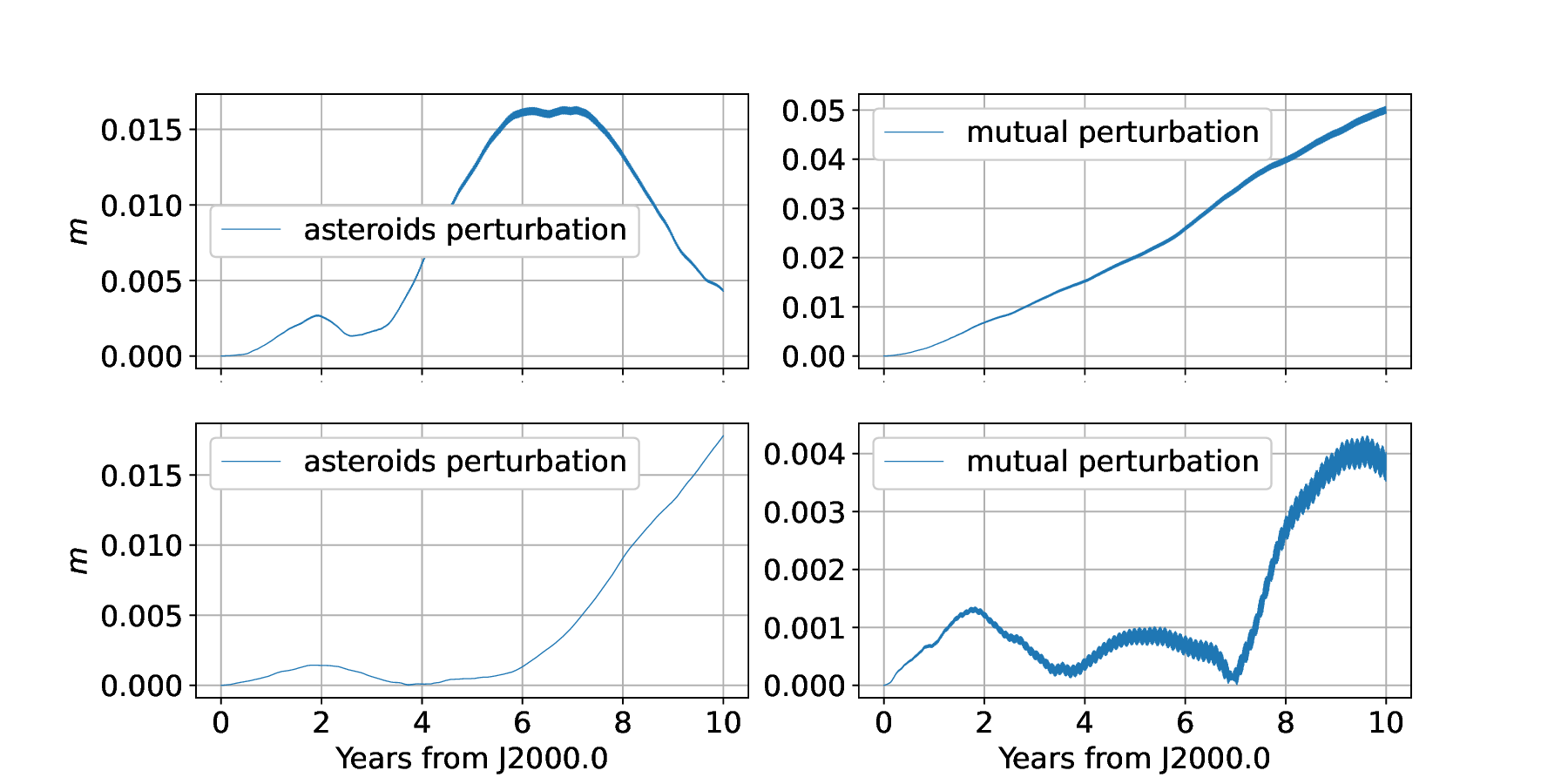}
   \caption{Orbital differences with and without asteroid perturbations and mutual perturbations. The perturbations tested here are the three largest asteroids acting on Phobos (upper left panel) and Deimos (lower left panel), and the mutual point mass torques on Phobos (upper right panel) and Deimos (lower right panel).}
   \label{Fig9}
   \end{figure}

\begin{figure} [!htbp]
   \centering
   \includegraphics[width=12.0cm, angle=0]{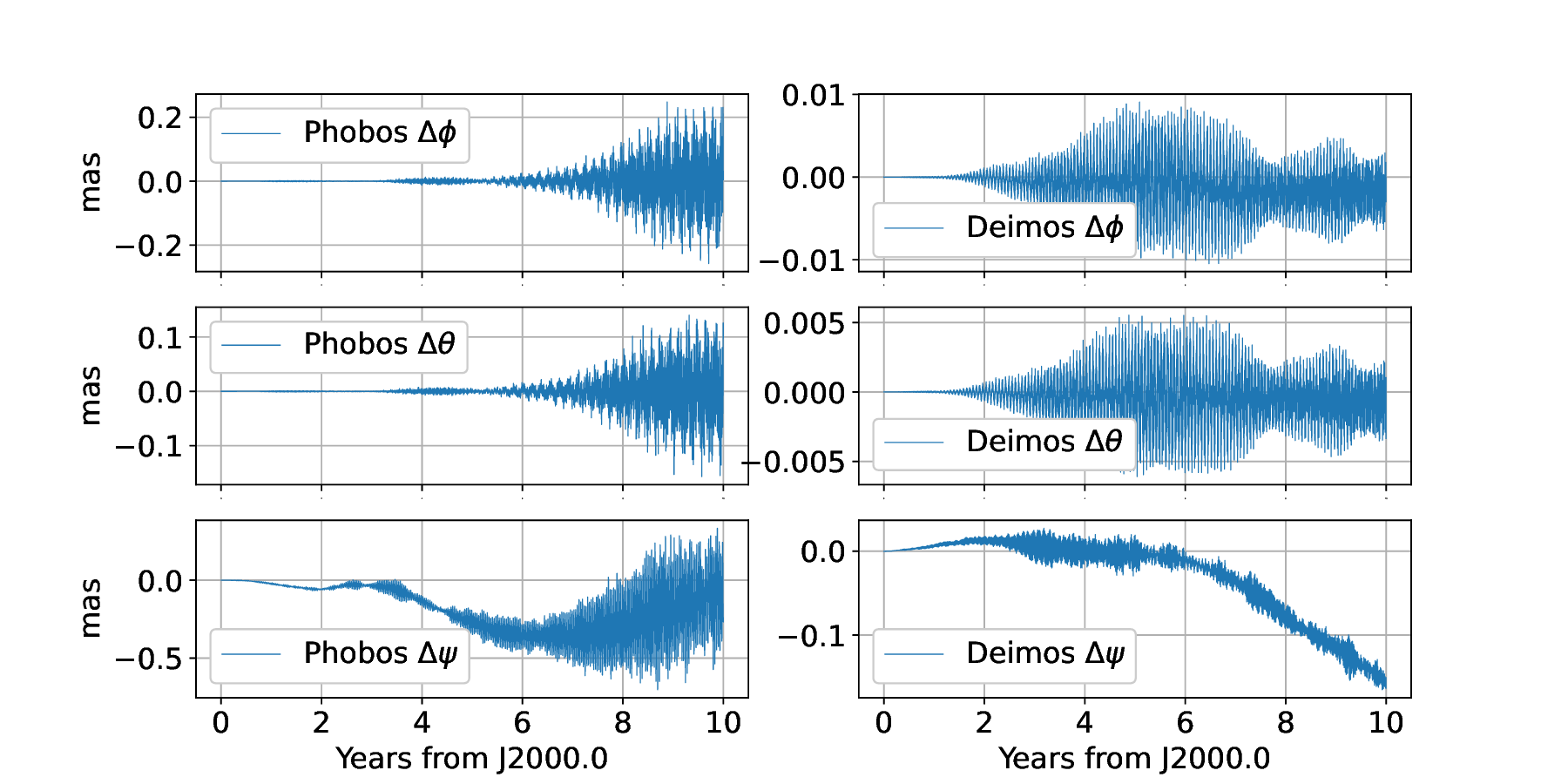}
   \caption{Euler angles differences between two cases involving asteroids' perturbations or not, left panels for Phobos and right for Deimos.} 
   \label{Fig10}
   \end{figure}

\begin{figure} [!htbp]
   \centering
   \includegraphics[width=12.0cm, angle=0]{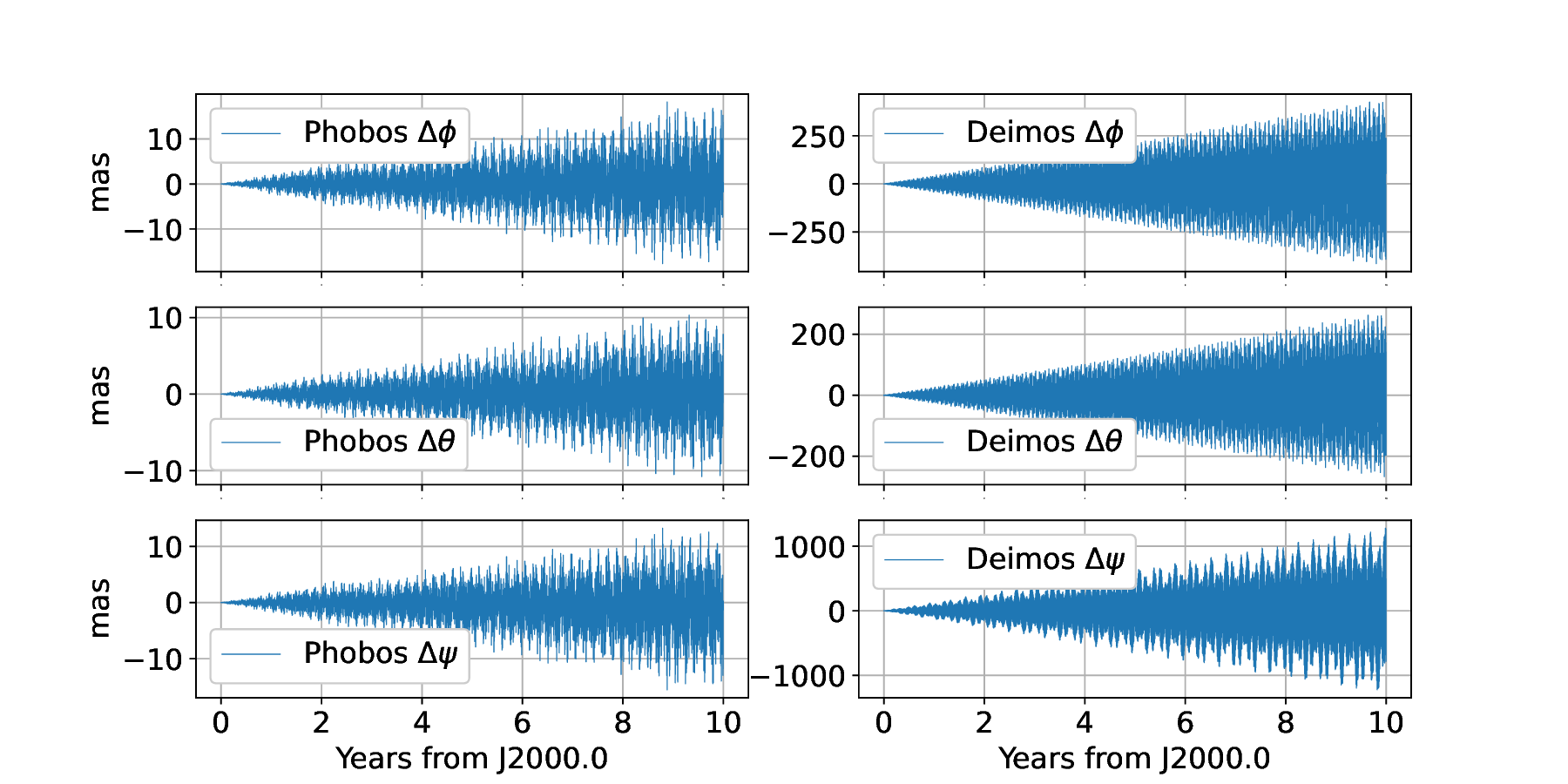}
   \caption{Euler angles differences between two cases involving mutual point mass torque or not, left panels for Phobos and right for Deimos.} 
   \label{Fig11}
   \end{figure}

With the fitted initial conditions, we test here the two perturbations that were not introduced in the newly established dynamical model. An easy way to quantify them is to perform the difference between a first simulation involving the perturbation and a second simulation without them. The differences between one simulation with and one simulation without each perturbation tested are integrated over 10 years and presented in Figs.~\ref{Fig9} -~\ref{Fig11}. The first perturbation tested here is the influence of the three largest asteroids, 1 Ceres, 2 Pallas, and 4 Vesta. Their orbit informations are taken from the JPL SPICE kernel files (\url{https://naif.jpl.nasa.gov/pub/naif/}). The simulations indicate that this perturbation introduces only several centimeters of influence on the satellites' orbit, the effect on rotation is also very small, less than 0.2 milli-arcsecond. 

The second perturbation that has been tested is the presence of the mutual mass torques between the two martian moons.
For example, the point torque from Deimos to Phobos' figure can be calculated by: 
\begin{equation}
\begin{split}
\mathbf{T}_{figP-pmD} =  \bm{f}_{figP-pmD} \times (\bf{r_p - r_d}) \,,
\end{split}
\label{eq:mutp}
\end{equation}
where $\bm{f}_{figP-pmD}$ is the force acting on the Deimos as a point-mass in Phobos’ gravitational field, in turn one can calculate the torque of the point mass Phobos on Deimos.
The main related effect is that the torque on Deimos from Phobos’ point mass can lead to the difference in rotation angle reach up to 1 arc-second over 10 years. The results show that none of the two perturbations considered in this section appears to have an effect on the orbit at an observable level. For the rotation, the effect of the point mass torque by Phobos on Deimos' rotation can reach up to 1 arc-second, so this factor is recommended to be taken into account in the modeling process.
  
\section{Conclusion}
\label{sect:conclusion}

High-precision numerical ephemerides typically provide information on the position, velocity and orientation parameters of celestial bodies over time, in addition to allowing detailed studies of the evolution and internal structure of those bodies. This work developed a new numerical dynamical model of the motions of the Martian satellites, taking in full account their rotation, and constructed a dynamical model of the coupled orbits and rotations using a method often used in the study of the motion of the Moon \citep{pavlov2016determining, folkner2014planetary}. In order to study the differences between the newly developed model and the dynamical model now in use, we first reproduced the ephemerides model, which was fitted to the ephemerides NOE-4-2020 published by Paris Observatory \citep{lainey2020mars} with the least squares method, and then used it as the reference for which the newly developed model was fitted. The differences between the post-fit orbits of Phobos and Deimos for the two models are no more than 300~$m$ and 125~$m$, respectively.

For the first time, we have computed simultaneously the Euler angles and their rates of Phobos and Deimos by numerical integration \citep{rambaux2012rotational},  and confirmed that the results are in good agreement with IAU values. Moreover, we simulated two possible perturbations which were not adopted in our refined model, and find that their effects on the orbits are completely negligible. As for the effect on rotation, we propose to consider the role of mutual attraction on rotation.

This revised numerical model of the motion of the Martian satellites provides potential opportunities for further study of the Martian satellites using high-precision observations from future missions such as MMX. In the future, not only the positions but also orientation parameters of the satellites can be derived from the refined dynamical model \citep{yang2024numerical}. 
Finally, our improved model of the dynamics of the Martian satellites employs a generalized approach that can be extended to systems beyond the Martian system, such as Saturn and Jupiter, by appropriately treating the rotational model.

\normalem
\begin{acknowledgements}
This research was supported by the National Key Research and Development Program of China (2021YFA0715101), the National Natural Science Foundation of China (12033009, 12103087), the Strategic Priority Research Program of the Chinese Academy of Sciences (XDA0350300), the International Partnership Program of Chinese Academy of Sciences (020GJHZ2022034FN), the Yunnan Fundamental Research Projects (202201AU070225, 202301AT070328, 202401AT070141), the Young Talent Project of Yunnan Revitalization Talent Support Program. We thank the two anonymous referees for their careful reviewing and nice suggestions that have been incorporated into and improved the paper. Y. Yang thanks Dr. Val{\'e}ry Lainey for his valuable suggestions which helped to improve the paper, thanks Dr. Qingbao He for his help on spectrum analysis. Martian moons' ephemerides files can be found at \url{ftp://ftp.imcce.fr/pub/ephem/NOE/}(NOE-4-2020). 

\end{acknowledgements}
\bibliography{bibtex}{}
\bibliographystyle{aasjournal}



\end{document}